\documentclass[pss,fleqn,embeddedheads]{w-art}
\usepackage{times}
\usepackage{w-thm}
\usepackage[]{graphicx}
\usepackage{subfigure}
\begin{document}
\DOIsuffix{theDOIsuffix}
\Volume{XX}
\Issue{1}
\Copyrightissue{01}
\Month{01}
\Year{2003}
\pagespan{1}{}
\Receiveddate{\sf } 
\Reviseddate{\sf  } 
\Accepteddate{\sf  } 
\Dateposted{\sf  }
\keywords{manganites, superexchange, spin-orbital model, 
          magnetic order, orbital order.}
\subjclass[pacs]{75.30.Et, 75.47.Lx, 71.27.+a}

\newcommand{\Eq}[1]{{Eq.~(\ref{#1})}}
\newcommand{\E}{{\textrm{e}}}

\title{Temperature effects in a spin-orbital model for manganites}

\author[Daghofer]{M.~Daghofer\footnote{Corresponding
     author: e-mail: {\sf daghofer@itp.tu-graz.ac.at}}\inst{1,2}} 
\author[Neuber]{D.~R.~Neuber\inst{1}}
\author[Ole\'{s}]{Andrzej M. Ole\'{s} \inst{2,3}}
\address[\inst{1}]{Institute of
   Theoretical and Computational Physics, Graz University of Technology,\\
   Petersgasse 16, A-8010 Graz, Austria} 
\address[\inst{2}]{Max-Planck-Institut f\"ur Festk\"orperforschung,
              Heisenbergstrasse 1, D-70569 Stuttgart, Germany} 
\address[\inst{3}]{M. Smoluchowski Institute of Physics, Jagellonian 
                   University, Reymonta 4, PL-30059 Krak\'ow, Poland}
\author[von der Linden]{Wolfgang von der Linden \inst{1}}
\begin{abstract}
We study a two-dimensional effective orbital superexchange model 
derived for strongly correlated $e_g$ electrons coupled to $t_{2g}$ core
spins in layered manganites. 
One finds that the ferromagnetic (FM) and antiferromagnetic (AF) 
correlations closely compete, and small changes of parameters can switch 
the type of magnetic order. For the same reason, spin order is easily 
destroyed with rising temperature, while alternating orbital 
correlations can persist to temperatures where FM order has already 
melted. A scenario for the AF phase observed in LaSrMnO$_4$ is presented.

Published in phys. stat. sol. (b) {\bf 243}, 277 (2006)
\end{abstract}

\maketitle                   
\section{Introduction}

Manganites are characterized by a complex interplay of charge, spin, 
orbital and lattice degrees of freedom. Undoped and weakly doped single 
layer manganites La$_{1-x}$Sr$_{1+x}$MnO$_4$ show antiferromagnetic (AF) 
order \cite{Sen04}, in contrast to three-dimensional (3D) cubic 
La$_{1-x}$Sr$_x$MnO$_3$ compounds \cite{Dag02}, where ferromagnetic (FM) 
$ab$ planes are stacked antiferromagnetically in the undoped case (one
electron per site, $x=0$). 
As for their 3D counterparts, the properties of the layered systems are 
strongly influenced by the orbital degrees of freedom of $e_g$ electrons 
which couple to the spins and thus influence the magnetic order.

For $x=0$, the hopping of $e_g$ electrons is blocked 
by large Coulomb interaction $U$, and charge fluctuations are quenched. 
They may be treated by second order perturbation theory which leads to 
virtual $d_i^4d_j^4\rightleftharpoons d_i^5d_j^3$ excitations by either 
$e_g$ or $t_{2g}$ electrons, generating the spin-orbital superexchange 
\cite{Fei99}. In the present paper we intend to focus on the role played 
by the orbital degrees of freedom of $e_g$ electrons in stabilizing 
various types of magnetic order in the undoped single layer manganites, 
and the changes of spin and orbital correlations with increasing 
temperature. 

\section{ Effective spin-orbital model at finite temperature }

We study an effective spin-orbital superexchange model derived for the 
one-dimensional (1D) chain \cite{Dag04}, generalized here to 
two-dimensional (2D) planes of layered manganites. 
The model depends on spin configuration ${\cal S}$ and takes the form,
\begin{equation}
{\cal H}({\cal S})=H_{J'}+H_J+H_z.
\label{HtJ}
\end{equation}
It consists of:  
($i$) the superexchange $H_{J'}$ for the core spins formed by $t_{2g}$ 
electrons, 
($ii$) the orbital superexchange $H_J$ for the $e_g$ electrons, and
($iii$) a crystal field term $H_z$. 
In the limit of large Hund's exchange interaction, we 
restrict the $e_g$ electron configurations to the subspace with their 
spins parallel to the core spins at each site, and thus arrive at 
spinless fermions with an orbital flavor $\alpha=x,z$, standing for
$|x\rangle\equiv\frac{1}{\sqrt{2}} |x^2-y^2\rangle$ and 
$|z\rangle\equiv\frac{1}{\sqrt{6}}|3z^2-r^2\rangle$ orbitals, 
respectively. Electrons with antiparallel spins are treated in second order perturbation
theory in the same way as double occupancies.

We start the discussion with the $t_{2g}$ core spin superexchange term 
$H_{J'}$. In reality it depends on the total number of $d$ electrons 
occupying two interacting Mn ions \cite{Ole02}. However, in the undoped 
systems one finds $d^4$ configuration at each site and one may replace 
this part of the superexchange by a Heisenberg Hamiltonian with a 
properly chosen exchange constant $J'>0$ favoring AF order:
\begin{equation}
\label{HJ'}
H_{J'} = J'\sum_{i}\big({\vec S}_i\cdot{\vec S}_{i+1}-S^2\big).
\end{equation}
For convenience we use classical core spins ${\vec S}_i$ of unit length, 
and compensate their physical value $S=3/2$ by a proper increase of $J'$. 
For classical spins ${\vec S}_i$ parameterized by polar angles
$\{\vartheta_i,\phi_i\}$, 
the spin product is given by 
$\langle {\vec S}_i\cdot{\vec S}_{j}\rangle=S^2\big(2|u_{ij}|^2-1\big)$, 
where the spin orientation enters via 
\begin{equation}
\label{uij}
u_{ij}=\cos(\vartheta_i/2)\cos(\vartheta_j/2) + 
\sin(\vartheta_i/2)\sin(\vartheta_j/2)\E^{i(\phi_j - \phi_i)}
=\cos(\theta_{ij}/2)e^{i\chi_{ij}}\;,
\end{equation}
depending on the angle $\theta_{ij}$ between the spins and on the phase 
$\chi_{ij}$, which cancels at $x=0$.

The superexchange term $H_J$ describes $e_g$ orbital interactions, as 
derived recently for a 1D chain \cite{Dag04} from the full coupled 
spin-orbital dynamics by replacing the spin scalar products by the 
actual averages $u_{ij}$. The orbital superexchange $H_J$ depends then 
on spin correlations via the factors $\{u_{ij}\}$ in Eq.~(\ref{uij}),
\begin{equation}
\label{HJ}
H_J= \frac{1}{5}J\sum_{\langle ij \rangle\parallel ab}
\big(2|u_{ij}|^2+3\big)\Big( 2T_{i}^{\zeta}T_{j}^{\zeta}
-\frac{1}{2}{\tilde n}_i{\tilde n}_{j} \Big) 
 - \frac{9}{10}J\sum_{\langle ij \rangle\parallel ab}
\big(1-|u_{ij}|^2\big){\tilde n}_{i\zeta}{\tilde n}_{j\zeta}.
\end{equation}
Here, it is most convenient to use directional orbitals along the $a$ and 
$b$ axes, with 
${\tilde c}_{i\zeta}^{\dagger} = -\frac{1}{2}{\tilde c}_{iz}^{\dagger}
\pm\frac{\sqrt{3}}{2}{\tilde c}_{ix}^{\dagger}$. The density operators:
${\tilde n}_{i\zeta}= {\tilde c}^\dagger_{i\zeta}{\tilde c}_{i\zeta}$ 
and ${\tilde n}_{i} ={\tilde n}_{i\zeta}+{\tilde n}_{i\xi}$, where 
$\langle\zeta|\xi\rangle=0$, are restricted to single occupancies, and
$T_{i}^{\zeta}=-\frac{1}{2}T_{i}^z\pm\frac{\sqrt{3}}{2}T_{i}^x$ also 
depend on the direction of the bond $\langle ij \rangle$. The pseudospin
$T=\frac{1}{2}$ orbital operators are given by
\begin{equation}
\label{TzTx}
T_{i}^{z}=\textstyle{\frac{1}{2}\sigma^z_i=\frac{1}{2}(n_{ix}-n_{iz})},\qquad  
T_{i}^{x}=\textstyle{\frac{1}{2}\sigma^x_i=\frac{1}{2}
(c_{ix}^{\dagger}c_{iz}^{}+c_{iz}^{\dagger}c_{ix}^{})}\;.
\end{equation}
Finally, the superexchange constant $J=t^2/\varepsilon(^6\!A_1)$ is given 
by the high-spin excitation energy $\varepsilon(^6\!A_1)$, see Ref.~\cite{Fei99}. 
The first term of Eq.(\ref{HJ}) favors FM spin and alternating orbital (AO) order, while 
the second one is optimized by AF spin and ferro orbital (FO) order --- 
both have similar strength and therefore closely compete.    

The last term of the Hamiltonian is a crystal field splitting of 
$e_g$ orbitals,
\begin{equation}
\label{Hz}
H_z=E_z\sum_i T_{i}^{z}
   =\frac{1}{2}E_z\sum_i ({\tilde n}_{ix}-{\tilde n}_{iz}), 
\end{equation}
with the orbital splitting $E_z$. We consider only $E_z>0$ with
out-of-plane $|z\rangle$ orbitals favored, which is suggested by the
elongated octahedra in LaSrMnO$_4$ along $c$ axis \cite{Sen04}.

In order to treat this model, we employ a Markov chain Monte Carlo (MCMC) 
algorithm for the classical core spins (see, e.g. Ref. \cite{Dag98}) 
combined with exact diagonalization for the electron degrees of freedom 
\cite{Dag04,Kol02}. In this method, the many-particle Hamiltonian 
for each spin configuration ${\cal S}$ determined by $\{u_{ij}\}$ is 
solved by exact diagonalization. The lowest states are then used to 
evaluate the trace over the fermionic degrees of freedom, 
$\textrm{Tr}_c\,\textrm{e}^{-\beta {\cal H(S)}}=:w({\cal S})$, which 
gives the statistical weight for a given spin configuration ${\cal S}$, 
and which is sampled by the MCMC. Autocorrelation analysis was used in 
order to verify that enough configurations were skipped between 
measurements. The Boltzmann factors of the Lanczos eigenstates are 
measured in order to ensure that only negligible weight is lost when the 
weight is calculated from the lowest eigenstates and we found that at most 
one or two percent are missed for $\beta t > 25$, where $\beta=1/k_BT$ is
inverse temperature. 

\section{ Numerical results and discussion }

\begin{figure}[t!]
  \subfigure{\includegraphics[height=0.26\textwidth]
    {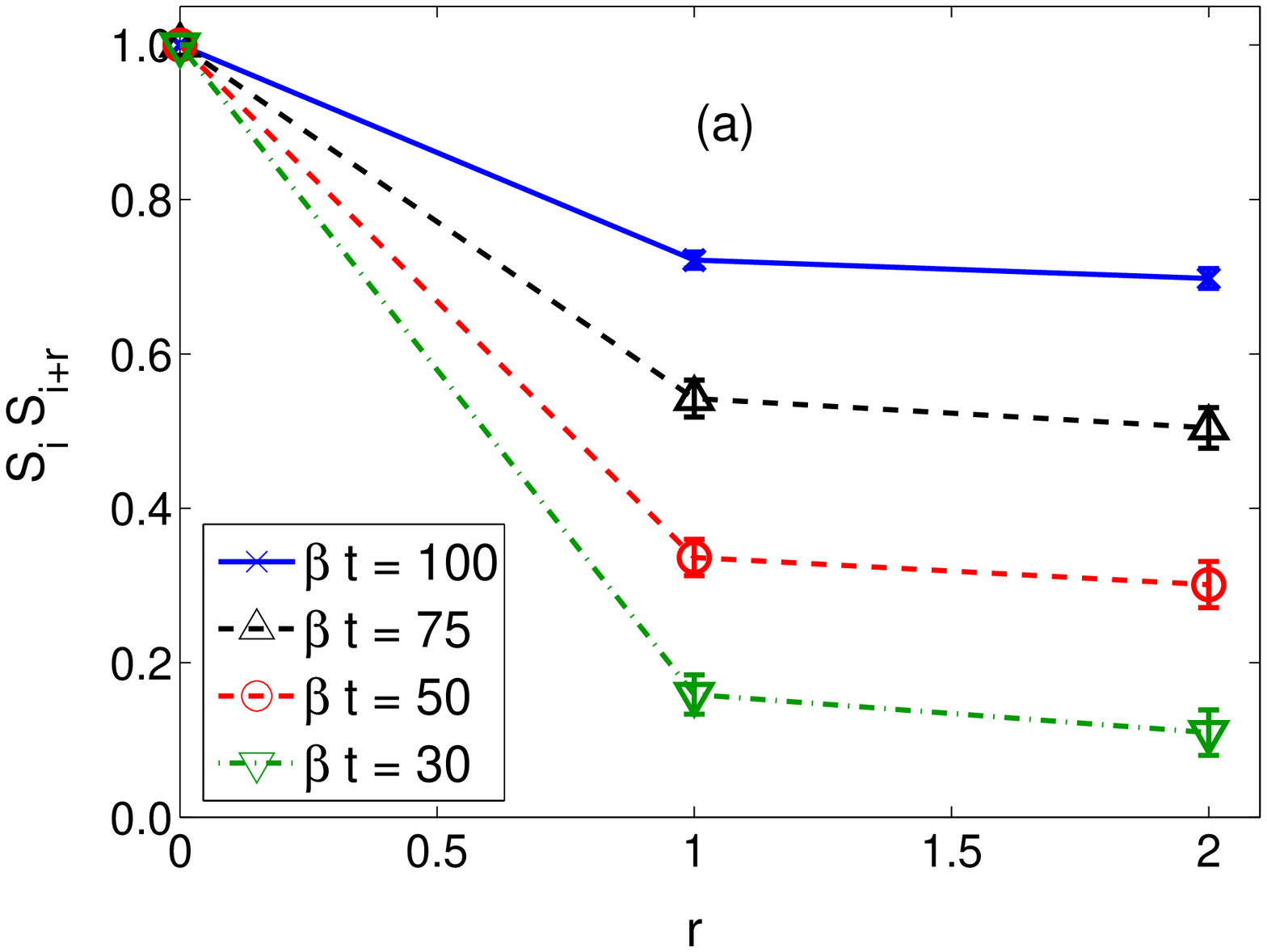}\label{fig:ss_J0_0h}}
  \subfigure{\includegraphics[height=0.26\textwidth]
    {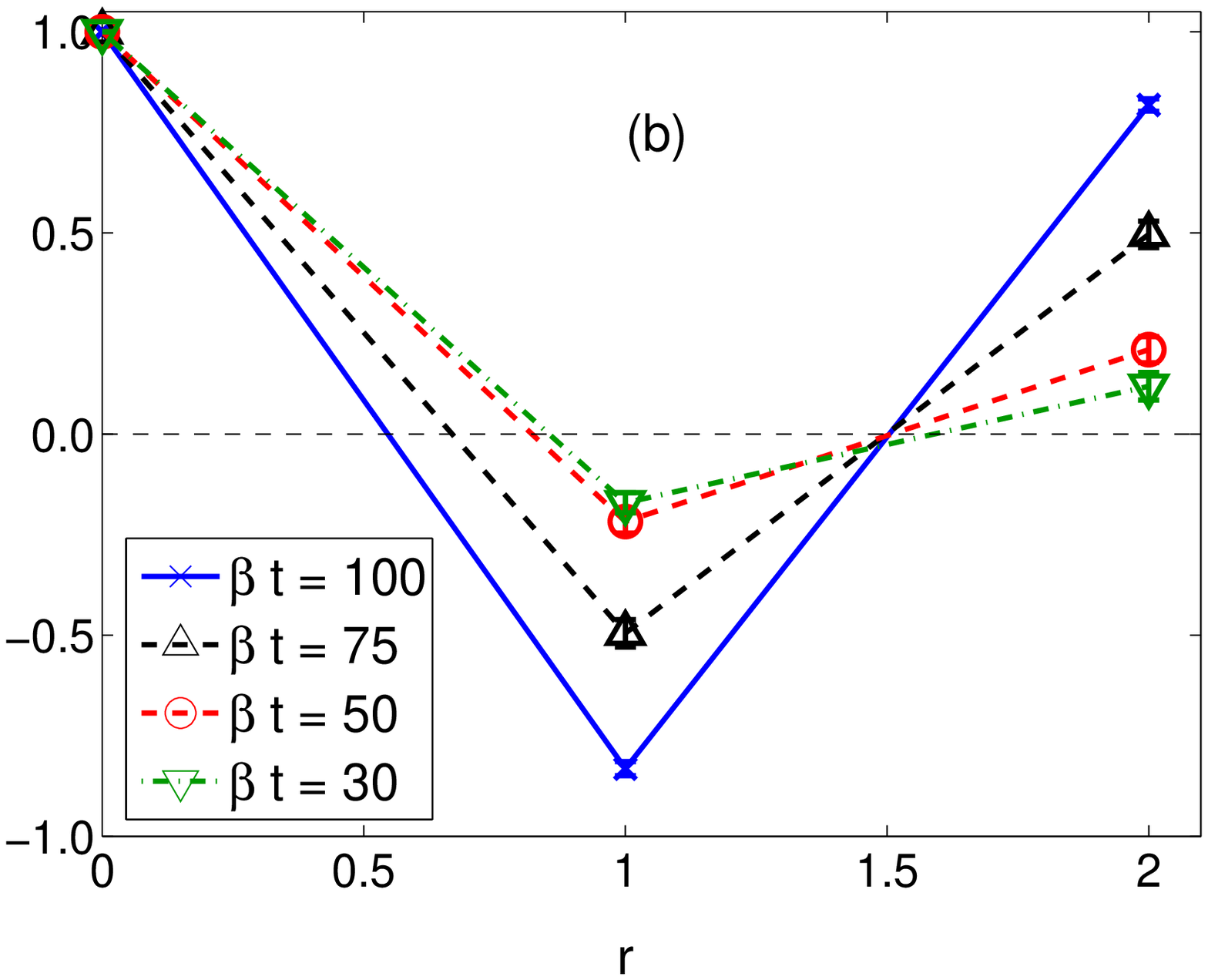}\label{fig:ss_J0.02_0h}}
  \subfigure{\includegraphics[height=0.26\textwidth]
    {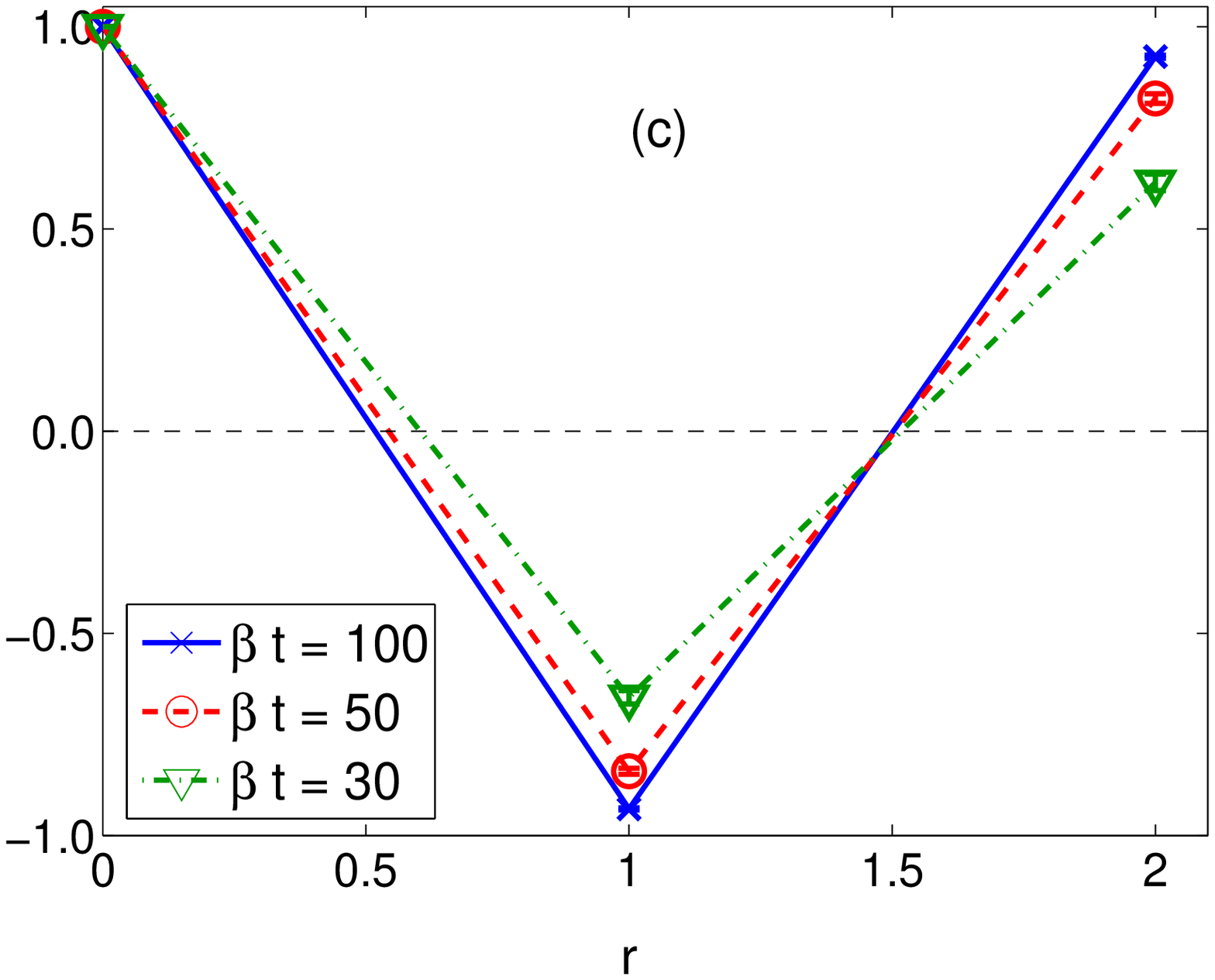}\label{fig:ss_J0.05_0h}}\\[-1em]
\caption{
Spin correlations along $(0,1)$-direction depending on distance $r$ at various temperatures, 
as obtained with $\sqrt{8}\times\sqrt{8}$ clusters in $ab$ plane for:
(a) $J'=0$,
(b) $J'=0.02t$, and 
(c) $J'=0.05t$. 
Parameters: $J/t=1/8$, and $E_z=0$.}
\label{fig:ss_0h} 
\end{figure}

Spin correlations $S(r)=\langle {\vec S}_i\cdot{\vec S}_{i+r}\rangle$
for the undoped system along the $(0,1)$-direction are gradually weakened
when temperature increases (Fig.~\ref{fig:ss_0h}). By increasing $J'$ one 
identifies two distinctly different cases: 
for $J'=0$ the system is FM at low temperature $\beta t=100$ 
[Fig.~\ref{fig:ss_J0_0h}], while 
 AF spin correlations arise for $J'=0.02t$ and $J'=0.05t$, leading to 
zig-zag lines in Figs. \ref{fig:ss_J0.02_0h} and \ref{fig:ss_J0.05_0h}. 
Already a small value $J'=0.02t$ suffices to switch the spin order from FM 
to AF, because FM and AF terms strongly compete in Eq. (\ref{HJ}). 
For the same reason, increasing temperature can easily destroy the 
magnetic order and both FM and AF correlations decrease rapidly around 
$\beta t\sim 50$, except for the stronger AF superexchange $J'=0.05t$, 
with more robust AF correlations. 

\begin{figure}[t!]
  \subfigure{\includegraphics[height=0.255\textwidth]
    {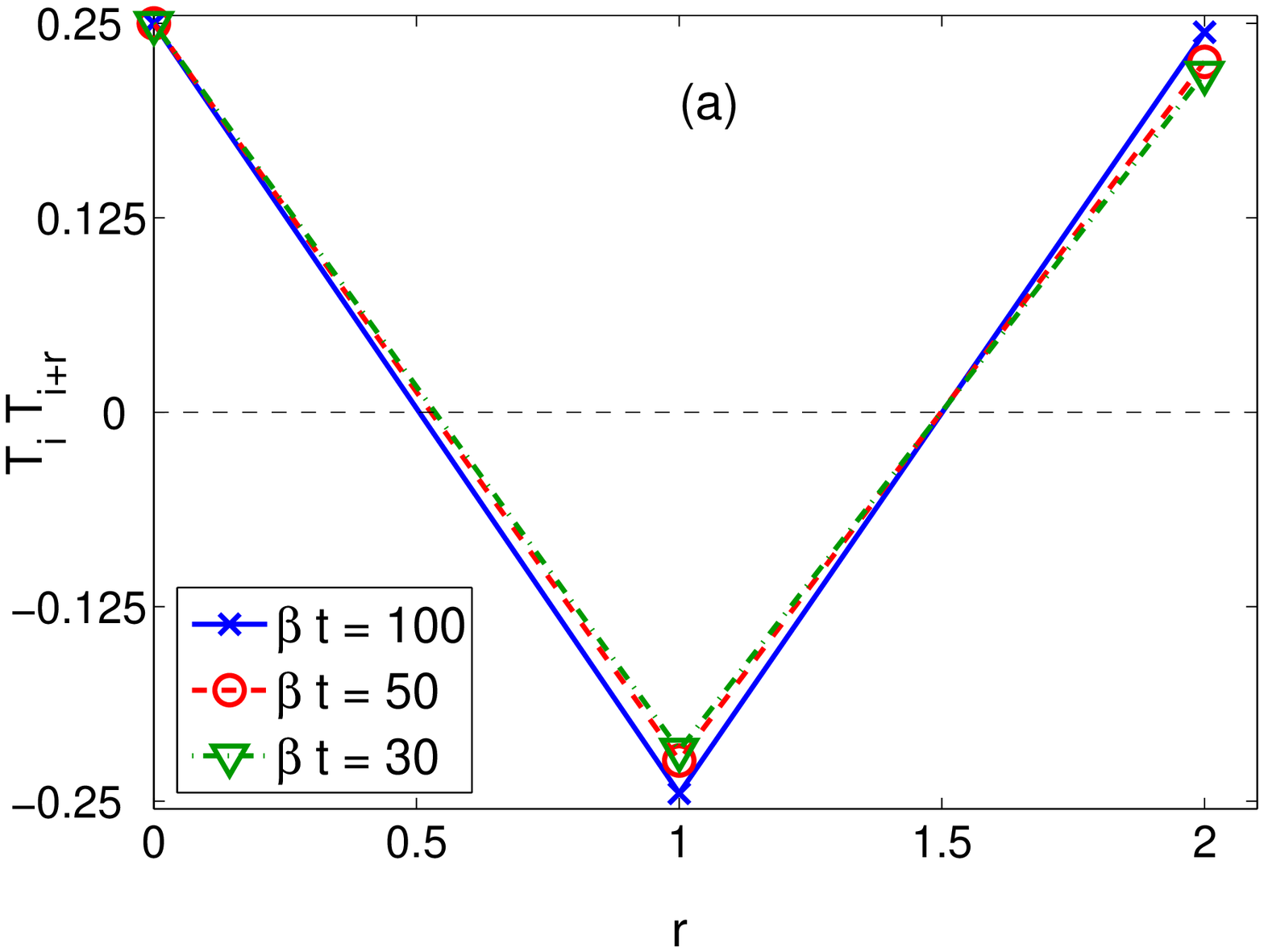}\label{fig:oo_J0_0h}}
  \subfigure{\includegraphics[height=0.255\textwidth]
    {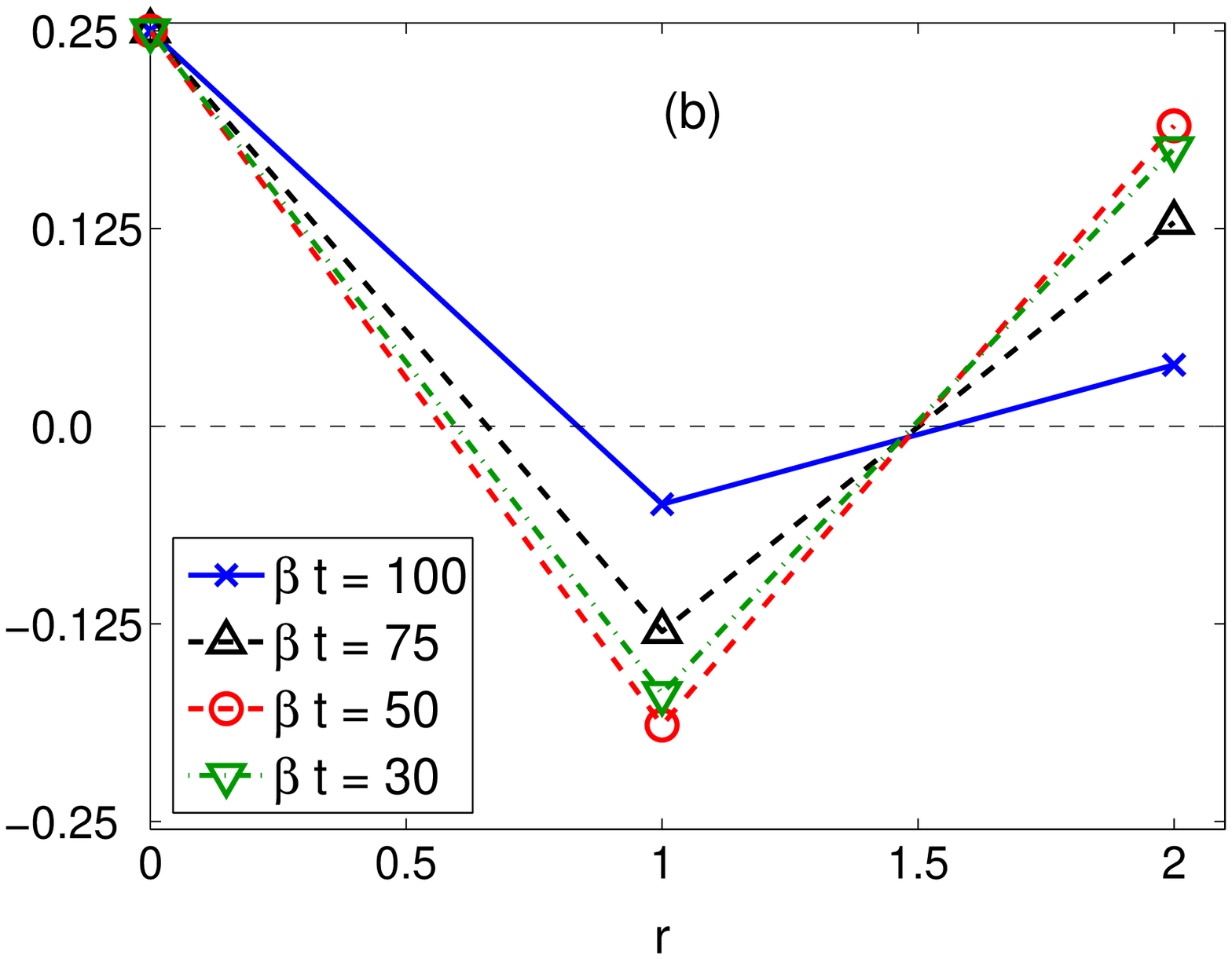}\label{fig:oo_J0.02_0h}}
  \subfigure{\includegraphics[height=0.255\textwidth]
    {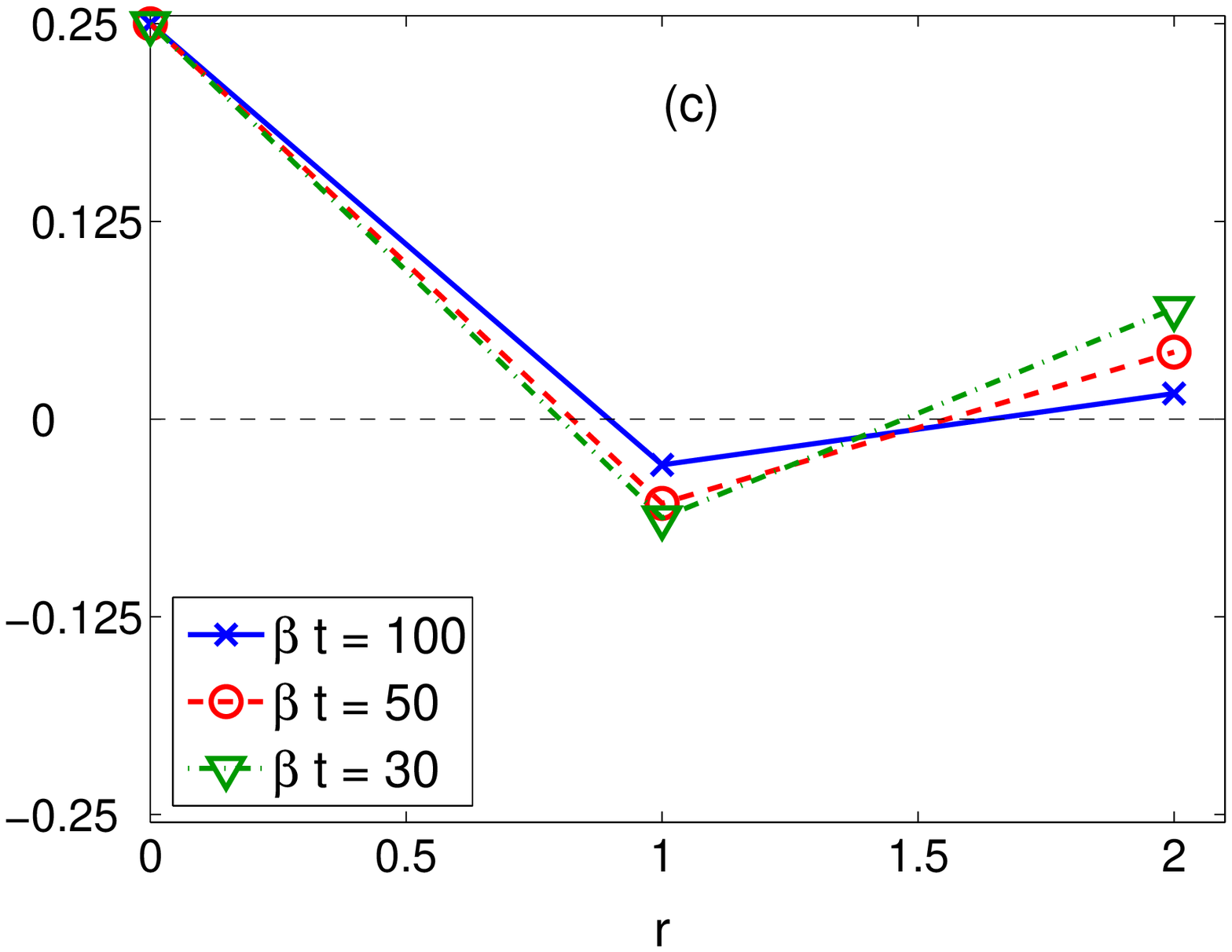}\label{fig:oo_J0.05_0h}}\\[-1em]
\caption{
Orbital correlation along $(0,1)$-direction depending on distance $r$ at various temperatures, 
as obtained with $\sqrt{8}\times\sqrt{8}$ clusters in $ab$ plane.
Parameters and temperatures as Fig.~\ref{fig:ss_0h}.}
\label{fig:oo_0h} 
\end{figure}

Next we investigate orbital correlations $T(r)=\langle T_iT_{i+r}\rangle$ 
shown in Fig.~\ref{fig:oo_0h}, which exhibit complementary behavior to the spin 
sector. Here we selected $T_i=\frac{1}{\sqrt{2}}(T^z_i+T^x_i)$ operators 
which correspond to the expected AO order with
$\frac{1}{\sqrt{2}}(|x\rangle\pm|z\rangle)$ orbital states in the 2D $ab$ 
plane. One finds that the orbital correlations alternate for the FM spin 
order, while this alternation is suppressed for increasing $J'$, as almost 
only orbitals within the plane are then occupied (FO order) which promotes 
the AF spin order. For $J'=0$, see Fig.~\ref{fig:oo_J0_0h}, one notes that 
the AO order remains robust with rising temperature although the spin 
correlations vanish, see Fig.~\ref{fig:ss_J0_0h}. Remarkably, in the AF 
case one observes the gradual development of AO order as the AF spin 
correlations become weaker. For $J'=0.02t$ this crossover happens at 
fairly low temperature $\beta t=50$, while it moves to higher temperature 
for $J'=0.05 t$.

\begin{figure}[t!]
\subfigure{\includegraphics[width=0.47\textwidth]{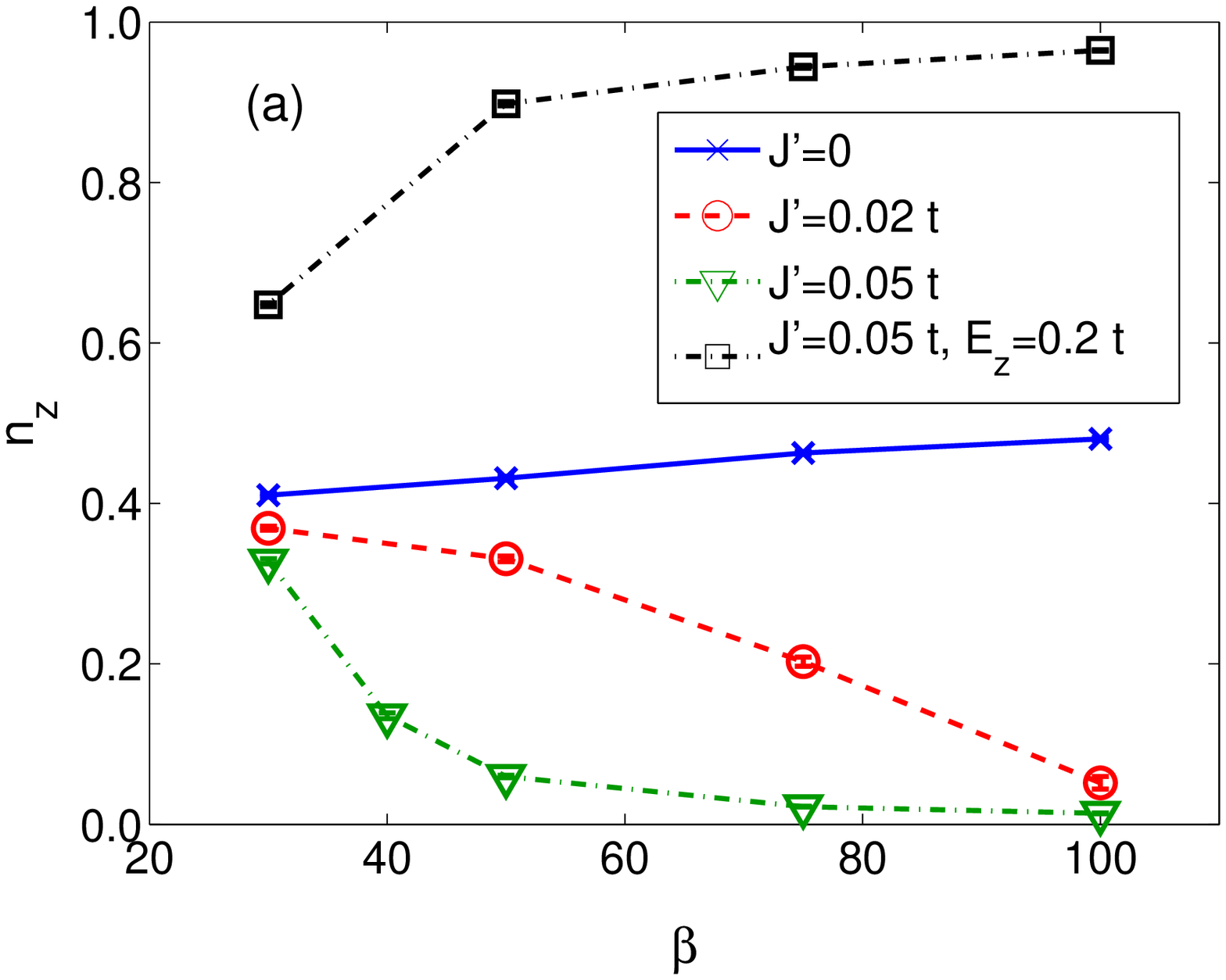}
\label{fig:n_beta}}\hspace{0.5cm}
\subfigure{\includegraphics[width=0.47\textwidth]{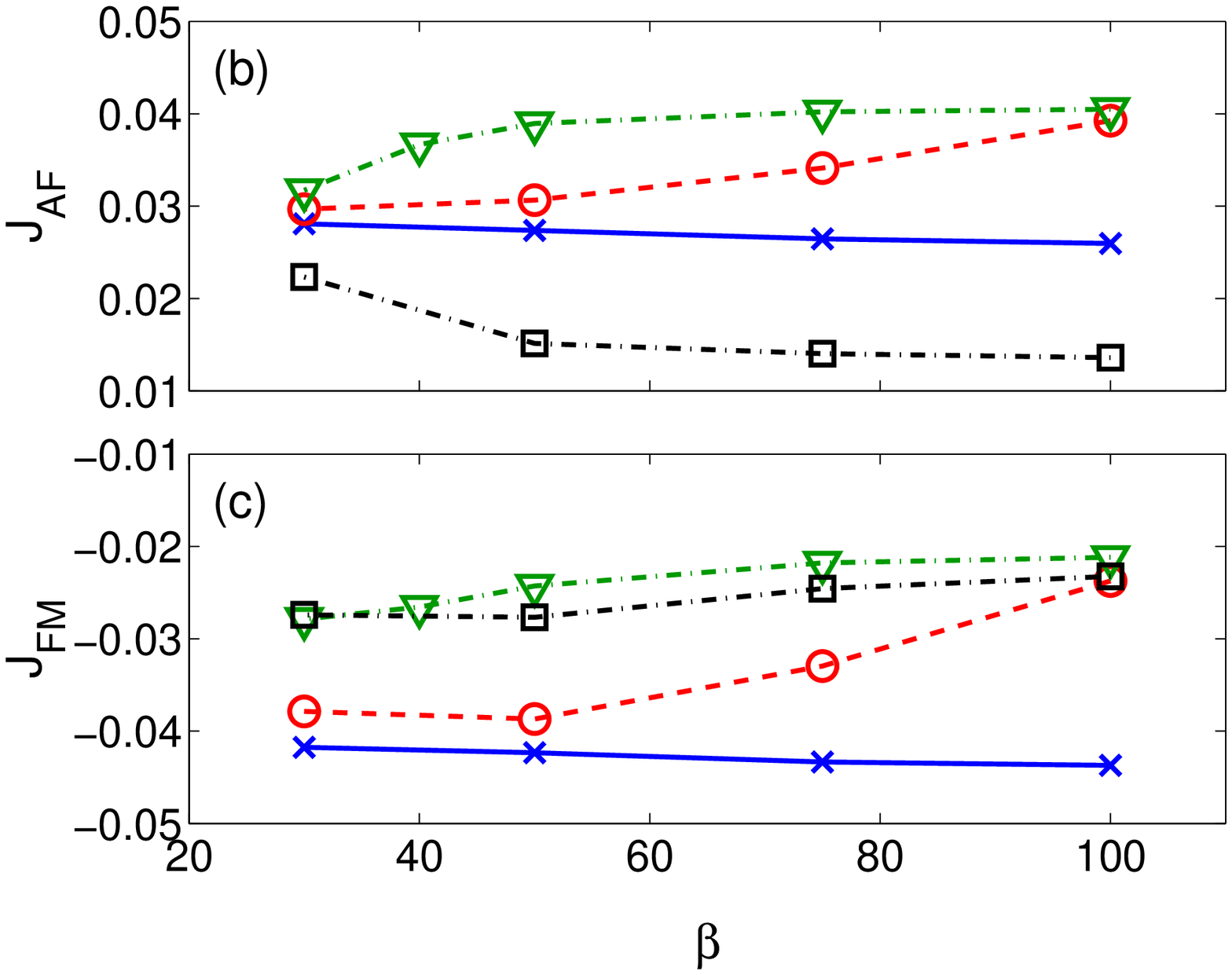}
\label{fig:se_fm_af}}
\caption{
Temperature dependence of:
(a) density in the out-of-plane $|z\rangle$ orbitals, as well as 
    average $e_g$ superexchange terms: 
(b) $J_{\rm AF}$ and 
(c) $J_{\rm FM}$. Remaining parameters as Fig.~\ref{fig:ss_J0.05_0h}.
\label{fig:n_se_beta}} 
\end{figure}

As the magnetic order is so closely related to the orbital order in $ab$ 
planes, one expects that changing orbital occupation will modify the 
magnetic order. Therefore, we investigated the electron density within 
the out-of-plane $|z\rangle$ orbitals, 
$n_z=\langle {\tilde n}_{iz}\rangle$. Indeed, $n_z\simeq 0.5$ in the FM 
phase at low temperature ($J'=0$), consistent with AO order, while FO
order with $|x\rangle$ orbitals ($n_z\sim 0$) takes over for the strongly 
AF case ($J'=0.05 t$).
With rising temperature, $|z\rangle$ orbitals are slightly depleted in 
the ferromagnet as orbital alternation becomes weaker. On the contrary, 
their occupation grows in the AF case, when such alternation develops. 
This happens quite soon for $J'=0.02 t$, but takes much longer for
$J'=0.05 t$, where AF spin order also persists to higher temperature, 
see Fig.~\ref{fig:oo_J0.05_0h}. Further, if one adds a crystal field 
splitting $E_z=0.2 t$ in the AF phase with $J'=0.05 t$, this forces almost 
all electrons into the out-of-plane $|z\rangle$ orbitals, the situation
encountered in LaSrMnO$_4$. With rising temperature, however, electrons 
redistribute and in-plane $|x\rangle$ orbitals gradually fill in, see 
Fig.~\ref{fig:n_beta}, which was also observed in experiment \cite{Sen04}.

The average AF and FM terms in $e_g$ superexchange can be derived from 
Ref. \cite{Fei99} and are given by:
$J_{\rm AF}= \frac{9}{20}J\langle {\tilde n}_{i\zeta}{\tilde n}_{j\zeta}
  \rangle-\frac{1}{5}J \langle  2T_{i}^{\zeta}T_{j}^{\zeta}
  -\textstyle{\frac{1}{2}{\tilde n}_i{\tilde n}_{j}} \rangle$, and 
$J_{\rm FM}= \frac{2}{5}J\langle  2T_{i}^{\zeta}T_{j}^{\zeta}
  -\textstyle{\frac{1}{2}{\tilde n}_i{\tilde n}_{j}} \rangle$.
Although $J_{\rm FM}$ becomes slightly smaller at higher temperature in 
the FM phase at $J'=0$ [Fig. \ref{fig:n_se_beta}(b)], the changes are not 
very marked, because AO order persists. The FM coupling is therefore still 
almost the same as at low temperature, even if the magnetic order is lost 
because of thermal fluctuations. In the AF systems ($J'=0.02t$, $0.05t$), 
the FM terms are first quenched by FO order, but become more important 
with rising temperature while the AF ones are reduced and, as before, 
this happens much sooner for $J'=0.02t$ than for $J'=0.05t$. 
Interestingly, the AF $e_g$ couplings are then changed into FM ones with 
rising temperature [Fig. \ref{fig:n_se_beta}(c)].

Summarizing, at finite crystal field $E_z\sim 0.2t$, as in LaSrMnO$_4$, 
one observes that both $J_{\rm FM}$ and $J_{\rm AF}$ are weaker than for 
$E_z=0$. The reason is that pairs of alternating orbitals are missing
for FO order with $|z\rangle$ orbitals in $ab$ plane (small 
$J_{\rm FM}$), and the overlap between $|z\rangle$ orbitals is much 
smaller than between $|x\rangle$ orbitals (small $J_{\rm AF}$). 
Therefore, the AF order observed in LaSrMnO$_4$ is promoted mainly by $J'$. 

\begin{acknowledgement}
  This work has been supported by the Austrian Science Fund (FWF), 
  Project No.~P15834-PHY, and by the Polish State Committee 
  of Scientific Research (KBN), Project No.~1 P03B 068 26.
\end{acknowledgement}

\end{document}